\documentclass[prl,aps,amsmath,twocolumn,showpacs,superscriptaddress,
longbibliography]{revtex4-1}

\usepackage{amssymb}
\usepackage{graphicx}
\usepackage[pdftex,bookmarks,colorlinks,breaklinks]{hyperref}
\hypersetup{linkcolor=red,citecolor=blue,filecolor=dullmagenta,urlcolor=blue}
\usepackage{color}
\usepackage{soul}
\definecolor{indiagreen}{rgb}{0.07, 0.53, 0.03}

\newcommand{\hj}[1]{\textcolor{blue}{#1}}

\begin{document}

%====================================================================++

\title{Transport studies in three-terminal microwave graphs with orthogonal, unitary, and symplectic symmetry}

\author{A.~M. Mart\'inez-Arg\"uello}
\affiliation{Instituto de F\'isica, Benem\'erita Universidad Aut\'onoma de
Puebla, Apartado Postal J-48, 72570 Puebla, Pue., Mexico}

\author{A. Rehemanjiang}
\affiliation{Fachbereich Physik der Philipps-Universit\"at Marburg, D-35032
Marburg, Germany}

\author{M. Mart\'{\i}nez-Mares}
\affiliation{Departamento de F\'{\i}sica, Universidad Aut\'onoma
Metropolitana-Iztapalapa, Apartado Postal 55-534, 09340 Ciudad de M\'exico,
Mexico}

\author{J.~A. M\'endez-Berm\'udez}
\affiliation{Instituto de F\'isica, Benem\'erita Universidad Aut\'onoma de
Puebla, Apartado Postal J-48, 72570 Puebla, Pue., Mexico}

\author{H.-J. St\"ockmann}
\affiliation{Fachbereich Physik der Philipps-Universit\"at Marburg, D-35032
Marburg, Germany}

\author{U. Kuhl}
\affiliation{Universit\'e de Nice-Sophia Antipolis, Laboratoire de la Physique
de la Mati\`ere Condens\'ee, CNRS, Parc Valrose, 06108 Nice, France}

%=======================================================================++

\begin{abstract}
The Landauer-B\"uttiker formalism establishes an equivalence between the
electrical conduction through a device, e.\,g.~a quantum dot, and the
transmission. Guided by this analogy we perform transmission measurements
through three-port microwave graphs with orthogonal, unitary, and symplectic
symmetry thus mimicking three-terminal voltage drop devices. One of the ports is
placed as input and a second one as output, while a third port is used as a
probe. Analytical predictions show good agreement with the measurements in the
presence of orthogonal and unitary symmetries, provided that the absorption and
the influence of the coupling port are taken into account. The symplectic
symmetry is realized in specifically designed graphs mimicking spin 1/2 systems.
Again a good agreement between experiment and theory is found. For the
symplectic case the results are marginally sensitive to absorption and coupling
strength of the port, in contrast to the orthogonal and unitary case.
\end{abstract}


\maketitle

%==================================================================++

Wave transport and wave scattering phenomena have been of great interest in the
last decades, both from experimental and theoretical points of view (see for
instance Ref.~\cite{JPA:MG}). Apart from the intrinsic importance in the
complex scattering in a particular medium, the interest also comes from the
equivalence between physical systems belonging to completely different areas,
in which the dimensions of the systems may differ by several orders of
magnitude~\cite{MelloBook}. One of these equivalences occurs in mesoscopic
quantum systems, where the electrical conduction reduces to a scattering
problem through the Landauer-B\"uttiker
formalism~\cite{Buttiker1986,ButtikerIBM,Landauer}. Following this line,
classical analogies of quantum systems have been used as auxiliary tools to
understand the properties of the conductance of electronic devices in
two-terminal
configurations~\cite{BB1997,Schanze2001,Schanze2005,MMM2005,EFlores}. A plethora
of chaotic scattering experiments in presence of time reversal invariance (TRI) and no spin 1/2 have been
performed~\cite{Schanze2001,Schanze2005,EFlores,Doron1990,Rafael2003,Kuhl2005,
Barthelemy2005,Hemmady2005,Hul2005}, while very few experimental studies regarding
absence of TRI are
reported~\cite{Schanze2001,Schanze2005,So1995,Stoffregen1995}. Furthermore, due
to its intrinsic complexity, there are no scattering experiments up to now for systems with TRI and spin 1/2,  where the signatures of the symplectic ensemble are expected, though there is one study of the spectral statistics in Au nanoparticles obeying this symmetry~\cite{AuNano}. Moreover, very recently the appearance of a microwave experiment showing the signatures of the symplectic symmetry~\cite{Abdu2016,Abdu2018} for eigenvalue statistics has opened the possibility to study transport in the presence of this symmetry.

\begin{figure}
\includegraphics[width=\columnwidth]{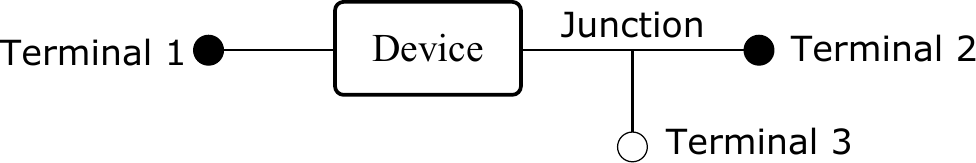}
\caption{\label{fig:buttiker3T}
Sketch of a three-terminal setting that allows the measurement of the voltage
along a device. The device carries a current while the vertical wire measures
the voltage drop. Thin lines represent perfect conductors connected to sources
of voltages $V_{1}$, $V_{2}$, and $V_{3}$.}
\end{figure}

Multiterminal devices are good candidates to provide experimental realizations
for the three symmetry classes: orthogonal, unitary and symplectic.
Alternatively to the most used two-terminal configuration, three terminal
systems provide information of nonlocal effects of transport observables. In the
present paper, we make theoretical predictions for coherent transport in a
three-terminal quantum device. In the spirit of the mentioned classical analogy,
we propose experimental realizations with microwave graphs, which represent the
first experiments of transport in three-terminal systems for the three
symmetry classes and the first experiment with the symplectic symmetry.

The electrical current $I_i$ on the terminal $i$ of an electronic device, as
given by B\"uttiker's formula, can be written as~\cite{DattaBook}
\begin{equation}
I_i = \sum_{j} G_{ij}(V_i-V_j), \quad\mbox{with}\quad
G_{ij} = \frac{e^2}{h} T_{ij},
\end{equation}
where $V_i$ is the voltage at terminal $i$, and $G_{ij}$ and $T_{ij}$ are the
conductance and transmission coefficient, respectively, from terminal $j$ to
terminal $i$. In a three-terminal configuration, one of the ports, let's say
terminal 3, can be used as a probe by tuning its voltage to zero current. This
voltage\, $V_3$, is a weighted average of the voltages in the other terminals,
the weight being determined by the conductance coefficients from the other
terminals to the probe~\cite{DattaBook}. It can be written
as~\cite{ButtikerIBM}
\begin{equation}
V_3 = \frac{1}{2}\left( V_1 + V_2\right) +
\frac{1}{2} \left( V_1 - V_2 \right)\, f,
\label{eq:mu3}
\end{equation}
where
\begin{equation}
f = \frac{T_{31} - T_{32}}{T_{31} + T_{32}},
\label{eq:f}
\end{equation}
see Fig.~\ref{fig:buttiker3T}. This equation shows that $V_3$ varies around the
average of the voltages producing the bias, $V_1$ and $V_2$. Hence, the quantity $f$ takes values in
the interval [-1, 1] and contains all the information about the system. For
instance, a three-terminal setting was considered in
Refs.~\cite{Godoy3T,Mello3T} to study the voltage drop along a disordered
quantum wire.

\begin{figure}
\begin{center}
\includegraphics[width=\columnwidth]{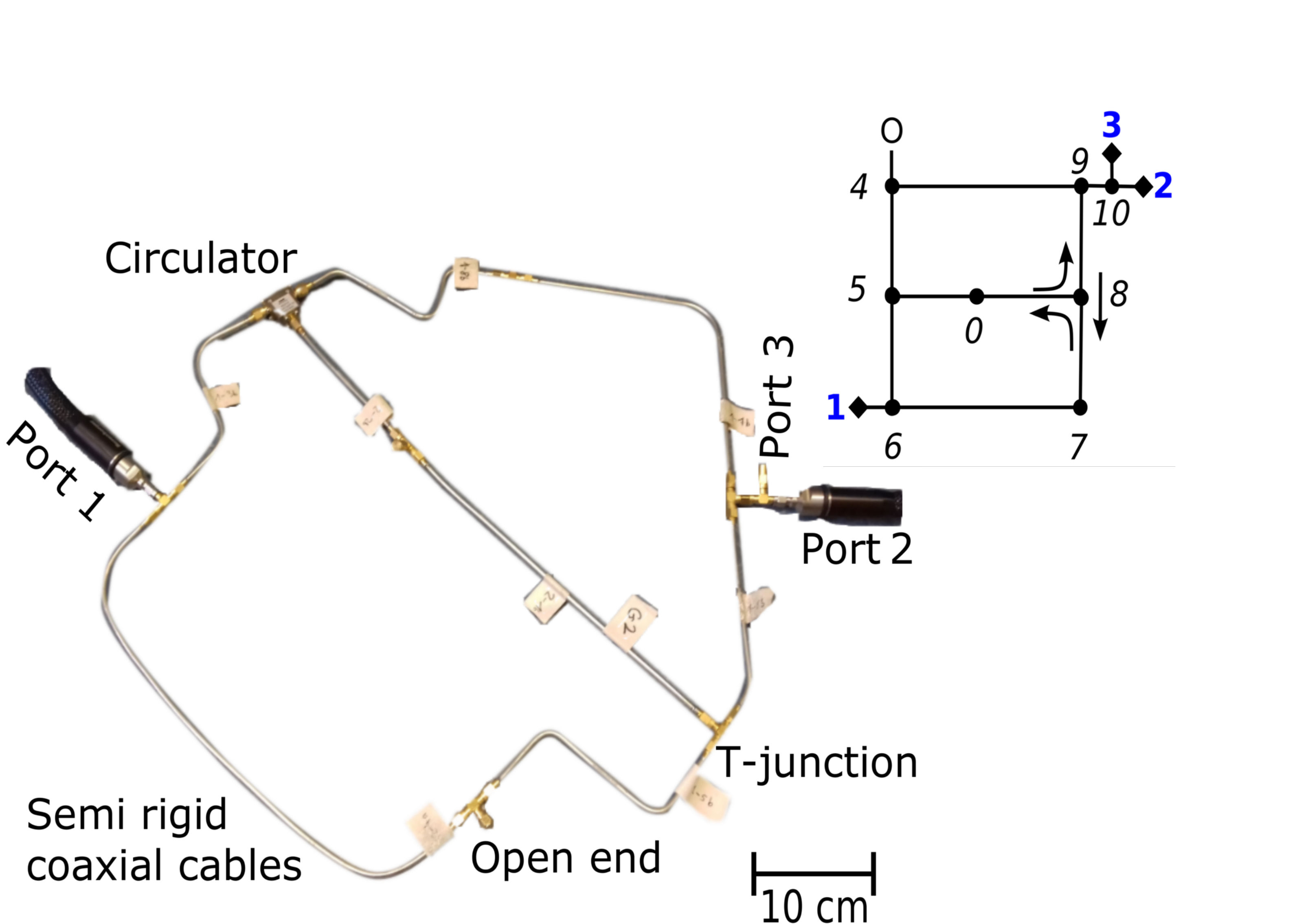}
\caption{\label{fig:GUEGraph}
Photograph of a three-port microwave graph with broken time reversal symmetry
(GUE). The circulator in the graph adds directionality, breaking TRI, and
yields to a GUE spectra. By replacing the circulator by an ordinary T-junction,
a graph with GOE spectra is obtained.}
\end{center}
\end{figure}

Here, we perform measurements of the quantity $f$ through microwave graphs
connected to three single channel ports: an input port, an output port, and a
probe port. We focus here on the particular situation where the probe port is on
one side of the microwave graph, see Fig.~\ref{fig:buttiker3T}. We study graphs
with chaotic dynamics characterized by the orthogonal, unitary, and symplectic
symmetries; labeled by $\beta=1$, 2, and 4 in Dyson's scheme~\cite{Dyson1962},
respectively. The $\beta=4$ case is realized in a network with specific
properties that mimics a spin 1/2 system \hj{\cite{Abdu2016,Abdu2018}}. The
fluctuations of $f$, that arise when the frequency is varied, are analyzed by
means of random matrix theory (RMT) calculations. Analytical expressions for the
distribution of $f$, that describe the experiments for the three symmetry
classes, are verified by the measurements.

\begin{figure}
\begin{center}
\includegraphics[width=\columnwidth]{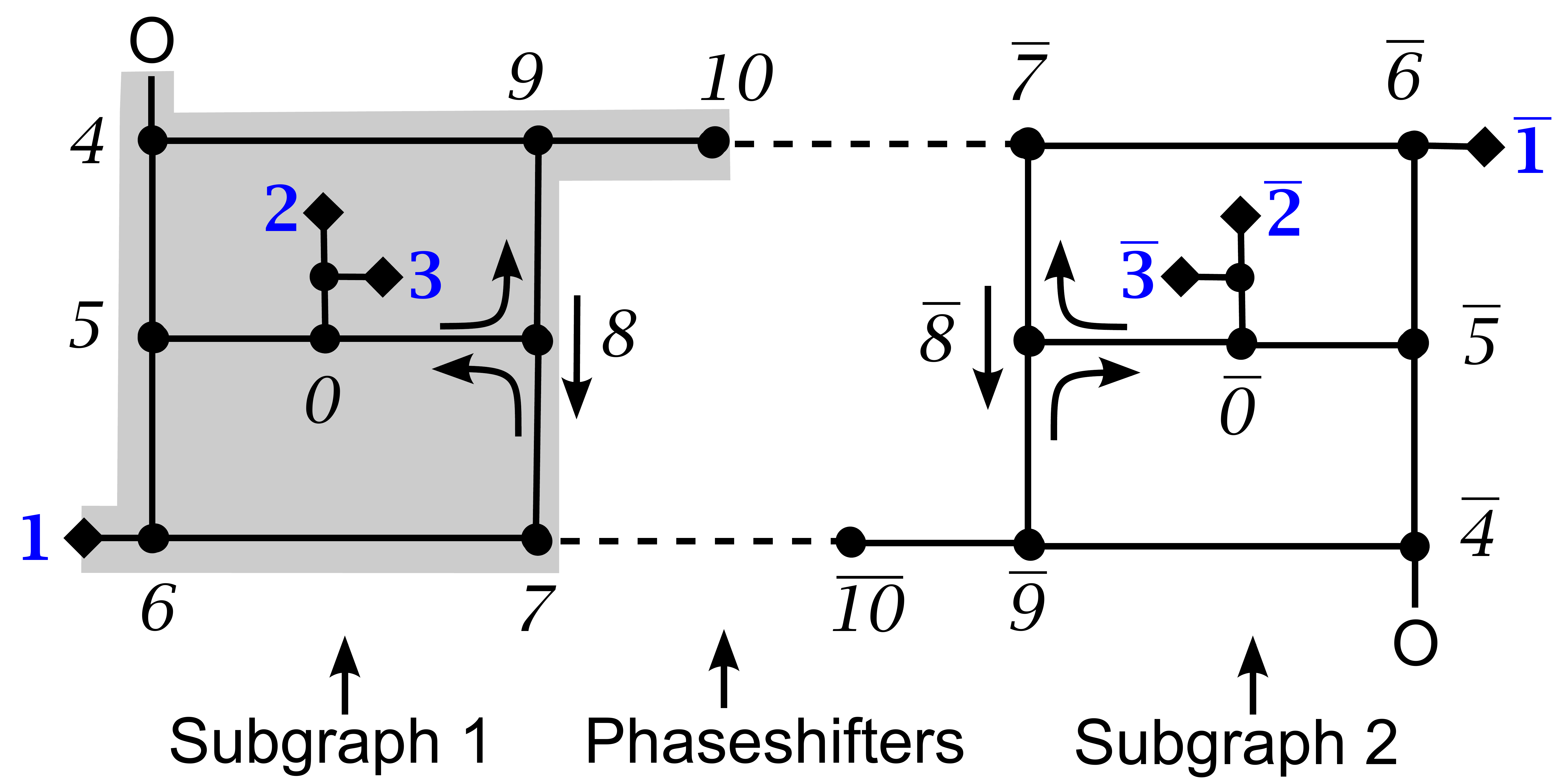}
\caption{\label{fig:GSEGraph}
Sketch of the three-terminal microwave GSE graph composed of two GUE
subgraphs. The circulators in the graphs add directionality in the system,
breaking TRI and yielding GUE spectra in each subgraph. Transmissions from ports
1 to 3, 1 to $\bar{3}$, $\bar{1}$ to 3, and from $\bar{1}$ to $\bar{3}$, are
measured to obtain $T_{31}$. In a similar way, $T_{32}$ is obtained.}
\end{center}
\end{figure}

The experimental setup for $\beta=2$ (with a small modification also for
$\beta=1$) is shown in Fig.~\ref{fig:GUEGraph}. A chaotic microwave graph is
formed by coaxial semirigid cables (Huber $\&$ Suhner EZ-141) with SMA
connectors, coupled by T-junctions at the nodes. An additional T-junction at
the exit port forms the three-terminal setting. For $\beta =1$ all bonds were
connected by T-junctions, for $\beta =2$ one of the T-junctions was replaced by
a circulator to break TRI. In both cases the found spectral level spacing
distributions were in perfect agreement with the Wigner distributions for the
Gaussian orthogonal ensemble (GOE), $\beta =1$, and the Gaussian unitary
ensemble (GUE), $\beta =2$, respectively, see e.\,g. Chapter 4.4 of
Ref.~\cite{haa01b}. The measurements were restricted to the operating range of
the circulators (Aerotek I70-1FFF) from 6 to 12\,GHz. To realize graphs showing
the signatures of the Gaussian symplectic ensemble (GSE), $\beta = 4$, two
copies of the graph shown in Fig.~\ref{fig:GUEGraph} are needed, where the
implemented circulators lead to an opposite sense of rotation. They are coupled
by two bonds in an inversion symmetric geometry, with a phase shift of $\pi$ in
one of the bonds but not the other one, see Fig.~\ref{fig:GSEGraph}. The whole
graph obeys an antiunitary symmetry $T$, squaring to -1, thus mimicking a spin
1/2, see Ref.~\cite{Abdu2016}. Transmission measurements were performed with an
Agilent 8720ES vector network analyzer (VNA).

With respect to the quantity $f$, its fluctuations can be described by the
scattering approach of RMT. Appealing to an ergodic hypothesis, fluctuations on
the frequency are replaced by fluctuations on an ensemble of chaotic graphs,
represented by an ensemble of scattering matrices. In the two-channel
situation, the scattering matrix of the graph has the structure
\begin{equation}
\label{eq:Sj}
S_{\mathrm{g}} =
\left( \begin{array}{cc}
r_{\mathrm{g}} & t'_{\mathrm{g}} \\
t_{\mathrm{g}} & r'_{\mathrm{g}} \\
\end{array} \right),
\end{equation}
where $r_{\mathrm{g}}$ ($r'_{\mathrm{g}}$) and $t_{\mathrm{g}}$
($t'_{\mathrm{g}}$) are the reflection and transmission amplitudes, for
incidence from the left (right). Depending on the symmetry class,
$S_{\mathrm{g}}$ belongs to one of the Circular Ensembles: Orthogonal (COE) for
$\beta=1$, Unitary (CUE) for $\beta=2$, and Symplectic (CSE) for $\beta=4$.
The $S_{\mathrm{g}}$ matrix can be written in the polar representation as~\cite{Brouwer1994}
\begin{equation}
\label{eq:Sjb1b2}
S_{\mathrm{g}} = \left[
\begin{array}{cc}
-\sqrt{1 - \tau}\, \mathrm{e}^{2\mathrm{i}\phi'} &
a^{-1}\sqrt{\tau}\, \mathrm{e}^{\mathrm{i}(\phi + \phi')} \\
a \sqrt{\tau}\, \mathrm{e}^{\mathrm{i}(\phi + \phi')} &
\sqrt{1 -\tau}\, \mathrm{e}^{2\mathrm{i}\phi}
\end{array}
\right],
\end{equation}
where $0\leqslant\tau\leqslant 1$, $0\leqslant\phi,\,\phi'\leqslant\pi$, and
$a$ is a real, complex, or real quaternion number of modulus 1 for $\beta=1$,
2 or 4, respectively. The probability density distribution of
$S_{\mathrm{g}}$ is given by~\cite{Brouwer1994}
\begin{equation}
\label{eq:JDP}
\mathrm{d}P_{\beta}(S_{\mathrm{g}}) =
\frac{\beta}{2}\, \tau^{-1+\beta/2}\,
\mathrm{d}\tau
\frac{\mathrm{d}\phi}{\pi}
\frac{\mathrm{d}\phi'}{\pi} \mathrm{d}a .
\end{equation}

The scattering matrix associated to the three-terminal setup of
Fig.~\ref{fig:buttiker3T}, where the probe is at the right of the graph, is
given by~\cite{Angel2014}
\begin{equation}
\label{eq:S3t}
S = S_{PP} +
S_{PQ} S_{0} \frac{1}{\openone_{3} - S_{QQ} S_{0}} S_{QP},
\end{equation}
where $S_{0}$ is the scattering matrix for the junction (see
Fig.~\ref{fig:buttiker3T}), $\openone_{3}$ stands for the unit matrix of
dimension 3, and
\begin{equation}
S_{PP} = \left(
\begin{array}{ccc}
r_{\mathrm{g}} & 0 & 0 \\
0 & 0 & 0 \\
0 & 0 & 0
\end{array}
\right) \label{eq:PP},
\quad
S_{PQ} = \left(
\begin{array}{ccc}
t'_{\mathrm{g}} & 0 & 0 \\
0 & 1 & 0 \\
0 & 0 & 1
\end{array}
\right),
\end{equation}
\begin{equation}
S_{QP} = \left(
\begin{array}{ccc}
t_{\mathrm{g}} & 0 & 0 \\
0 & 1 & 0 \\
0 & 0 & 1
\end{array}
\right) \label{eq:QP},
\quad
S_{QQ} = \left(
\begin{array}{ccc}
r'_{\mathrm{g}} & 0 & 0 \\
0 & 0 & 0 \\
0 & 0 & 0
\end{array}
\right).
\end{equation}
Equation (\ref{eq:S3t}) is a general combination rule for scattering matrices
which appears in several scattering problems. The first term, $S_{PP}$,
represents reflections on the terminals (only the Terminal 1 presents
reflection for the present case). The second term comes from multiple
scattering in the system. Reading from right to left, $S_{QP}$ represents the
transmissions to the inside region, passing through the graph and the
junction, $(\openone_3-S_{QQ}S_0)^{-1}$ contains the multiple reflections
between the junction and the graph, and $S_{PQ}$ gives the transmissions from
the internal region to the terminals.

Because it is expected that the T-junction couples the terminals
symmetrically, $S_0$ can be assumed to be symmetric. According to some
measurements~\cite{Allgaier2014}, it can be proposed as
\begin{equation}
S_{0} = \frac{1}{3}\left(
\begin{array}{ccc}
-1 & 2 & 2 \\
2 & -1 & 2 \\
2 & 2 & -1
\end{array}
\right) \label{eq:S0}.
\end{equation}
The general structure of $S$ is of the form
\begin{equation}
\label{eq:Sqij}
S = \left(
\begin{array}{ccc}
q_{_{1 1}} & q_{_{1 2}} & q_{_{1 3}} \\
q_{_{2 1}} & q_{_{2 2}} & q_{_{2 3}} \\
q_{_{3 1}} & q_{_{3 2}} & q_{_{3 3}}
\end{array}
\right),
\end{equation}
where $q_{_{ij}}=S_{ij}$ for $\beta=1$ and 2, while for $\beta=4$
\begin{equation}
\label{eq:qij}
q_{_{ij}} = \left(
\begin{array}{cc}
S_{i j} & S_{i \bar{j}} \\
S_{\bar{i} j} & S_{\bar{i}\,\bar{j}}
\end{array}
\right),
\end{equation}
where the ``bar'' in the subscripts denotes the corresponding terminal in
the second GUE subgraph needed for the construction of the GSE graph (see
Fig.~\ref{fig:GSEGraph}). Therefore, the transmission coefficient from
terminal $j$ to terminal $i$ is given by $T_{ij}=|S_{ij}|^2$ for $\beta=1$ and
2, and $T_{ij}=\frac{1}{2} \mathrm{tr}(q_{_{ij}} q^{\dagger}_{_{ij}})$ for
$\beta=4$.

Since $q_{_{ij}}$ is a quaternion real number, $q_{_{ij}}q^{\dagger}_{_{ij}}$
is proportional to the $2\times 2$ unit matrix. However, in the experiment
this can not be achieved with arbitrary accuracy due to power losses. For the
transmission measurements relevant to our study, they were realized within a
$10\%$ and $1\%$ of error for $q_{31}$ and $q_{32}$, respectively.

By substituting the parametrization given in Eq.~(\ref{eq:Sjb1b2}) into
Eqs.~(\ref{eq:S3t}) to (\ref{eq:QP}), and extracting the transmission
coefficients $T_{31}$ and $T_{32}$ from Eq.~(\ref{eq:Sqij}),
Eq.~(\ref{eq:f}) yields
\begin{equation}
\label{eq:fparametrica2}
f = \frac{\tau - |1 + \sqrt{1-\tau}\,
\mathrm{e}^{2\mathrm{i}\phi} |^{2}}{\tau + |1 + \sqrt{1-\tau}\,
\mathrm{e}^{2\mathrm{i}\phi} |^{2}} ,
\end{equation}
where $a$ and $\phi'$ drop out.

Using the probability density distribution of Eq.~(\ref{eq:JDP}) the
distribution of $f$ is obtained once the integration over all
parameters is done; the result is
\begin{equation}
 p_{_{\beta}}(f) = \frac{(\beta - 1)!!}{\beta\, [\Gamma(\beta/2)]^{2}}\,
\frac{(1 - f)^{\beta/2}}{(1 + f)^{1 - \beta/2}} .
\label{eq:pfAall}
\end{equation}
This distribution dominates for negative $f$ values in agreement with the
physical intuition since the probe at the right of the graph is closer to port 2
(see Fig.~\ref{fig:buttiker3T}), making the transmission $T_{32}$ predominantly
larger than the transmission $T_{31}$. The width of the distribution is a
signature of the nonlocal effects in the measurement of the probe port.

\begin{figure}
\begin{center}
\includegraphics[width=\columnwidth]{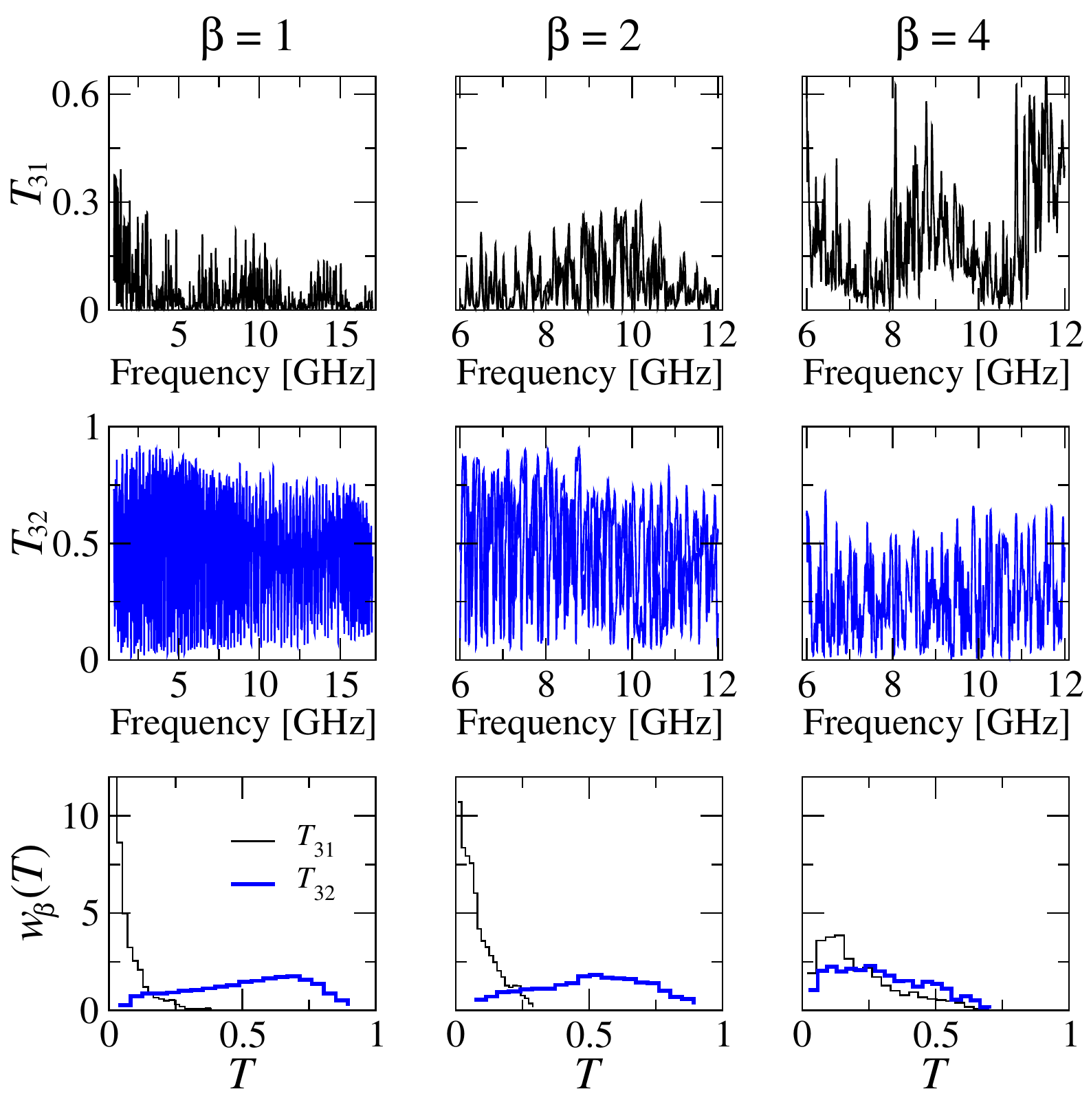}
\caption{{\footnotesize (color online) Transmissions, $T_{31}$ and $T_{32}$ as a function of frequency are shown in the upper and middle panels, and their corresponding distribution in thin black (thick blue) in the lower ones, for $\beta=1$ (left), 2 (middle), and 4 (right), respectively. }}
\label{fig:TpTAllB}
\end{center}
\end{figure}

Equation (\ref{eq:pfAall}) represents our main result which is valid in an
ideal situation: It applies to quantum systems in the absence of any
inelastic process and to classical wave systems in the absence of dissipation
and imperfect coupling to the device. In Fig.~\ref{fig:TpTAllB} we show the transmissions $T_{31}(= |S_{31}|^2)$ and $T_{32}(= |S_{31}|^2)$ as a function of the frequency, obtained from the measurements of the elements of the scattering matrix $S_{31}$ and $S_{32}$; for $\beta=1, 2$ and 4. We observe that they do not reach the value of 1 due to the losses of power. Their corresponding distribution are also shown.

\begin{figure*}
%\begin{figure}
\begin{center}
\includegraphics[width=14.0cm]{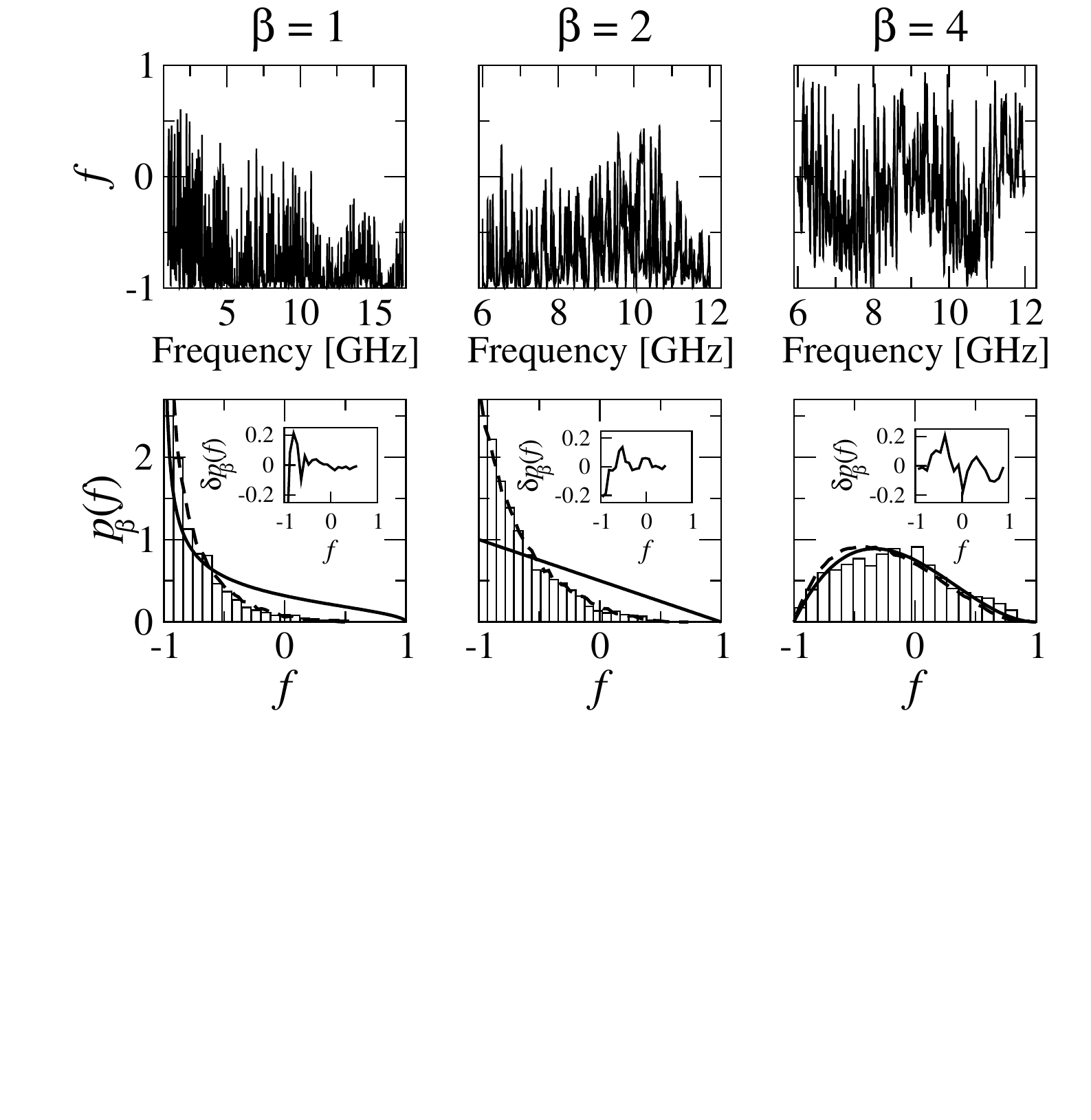}
\caption{{\footnotesize $f$ as a function of the frequency
is shown in the upper panels, and its corresponding distribution in the
lower ones, for $\beta=1$ (left), 2 (middle), and 4 (right). In the lower
panels the continuous lines represent the analytical result for the ideal case,
Eq.~(\ref{eq:pfAall}), while the dashed lines correspond to RMT simulations
with power losses and imperfect coupling of the T-junctions to the graph,
where all parameters were fixed before hand using the autocorrelation functions
(see Fig.~\ref{fig:Parameters2chAB1B2TA}). In the insets the difference between the numerical and the experimental distribution, $\delta p_{\beta}(f) = p_{\beta}(f)_{\mathrm{num}} - p_{\beta}(f)_{\mathrm{exp}}$, are presented for comparison purposes. For the statistical analysis we used an ensemble
of $5\times 10^4$ realizations.}}
\label{fig:pfAB1B2B4}
\end{center}
%\end{figure}
\end{figure*}

The actual measurements for $f$ (see Eq.~(\ref{eq:f})) are shown in
Fig.~\ref{fig:pfAB1B2B4} for experiments for the three symmetry classes:
$\beta=1$ (left panels), $\beta=2$ (middle panels), and $\beta=4$ (right
panels). In the upper panels we show the fluctuations of $f$ as a function of
the frequency for the three-terminal microwave setup of
Fig.~\ref{fig:buttiker3T}. For $\beta=1$, the spectrum was measured in the
frequency range from 1~GHz to 17~GHz, while for $\beta=2$ and 4 the interval
from 6~GHz to 12~GHz was considered due to the range of operation of the
circulators. In the lower panels the experimental distribution of $f$ is shown
as histograms. The solid lines correspond to the theoretical expectation, see
Eq.~(\ref{eq:pfAall}).

\begin{figure}
\begin{center}
\includegraphics[width=\columnwidth]{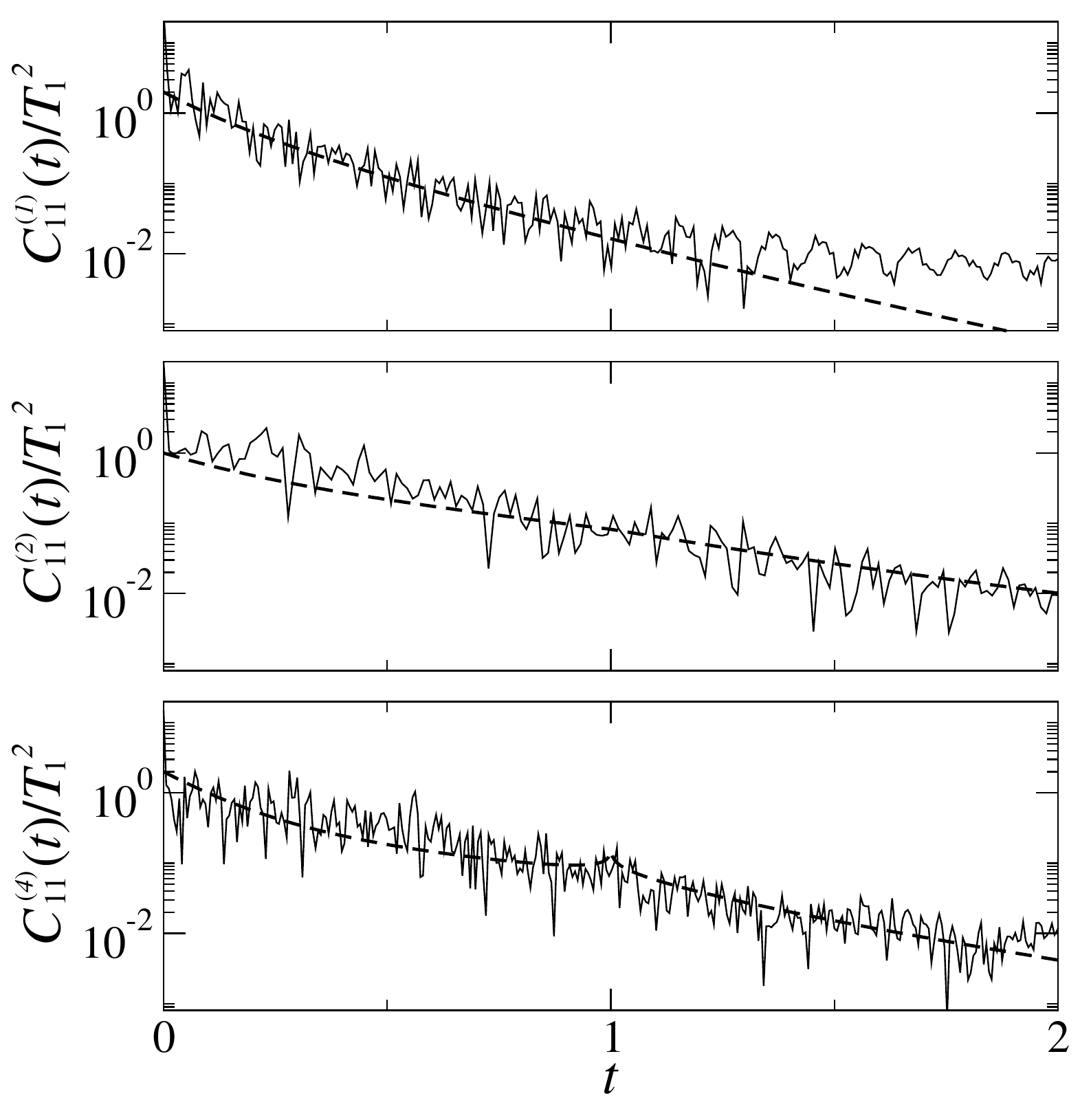}
\caption{{\footnotesize Fitting of the autocorrelation function,
Eq.~(\ref{eq:Autocorrelation}), to the experimental data. The parameters are
$T_{1}=0.98$ and $\gamma=1.9$ for $\beta=1$, $T_{1}=0.96$ and $\gamma=0.5$
for $\beta=2$, and $T_{1}=0.97$ and $\gamma=0.2$
for $\beta=4$. }}
\label{fig:Parameters2chAB1B2TA}
\end{center}
\end{figure}

The deviations between experiment and theory observed for the cases $\beta= 1$
and $2$ can be explained by the power losses and the imperfect coupling
between the graph and the ports. The effect of the absorption can be
quantified by assuming that the scattering matrix of the graph does not
conserve flux; while imperfect coupling can be modelled by adding identical
barriers, with transmission intensity $T_a$, between the graph and port 1,
between the graph and the T-junction, and between the T-junction and port 2,
respectively. Following Ref.~\cite{BB1997}, such scattering matrix, that we
denote by $\widetilde{S}_{\mathrm{g}}$, can be written as
\begin{equation}
\label{eq:Sabs}
\widetilde{S}_{\mathrm{g}}(E) = 1 - 2\pi \mathrm{i}
\widetilde{W}^{\dagger} \frac{1}{E - \widetilde{\mathcal{H}} +
\mathrm{i} \pi \widetilde{W} \widetilde{W}^{\dagger}} \widetilde{W} ,
\end{equation}
where $E$ is the energy and $\widetilde{W}$ accounts for the coupling between the resonant
modes of the graph and the scattering channels. Here,
$\widetilde{\mathcal{H}}_{mn}=H_{mn}+\mathrm{i}(\gamma\Delta/4\pi)\delta_{mn}$,
with $H$ being the Hamiltonian that describes the closed microwave graph with
mean level spacing $\Delta$ and it is taken from the Gaussian ensembles
corresponding to the symmetry present in the graph. The imaginary part of
$\widetilde{\mathcal{H}}$ mimics the absorption quantified by the parameter $\gamma$. It
can be extracted from the experimental data through the autocorrelation function, $C^{(\beta)}_{ab}(t)$,
between the elements of the scattering matrix $S_{ab}$. The corresponding expression for the
GOE is given in Ref.~\onlinecite{Schafer2003}, while for all $\beta$ in
Ref.~\onlinecite{Fyodorov2005}. After some mathematics, they can be written for the element $S_{11}$ as
\begin{equation}
\label{eq:Autocorrelation}
\frac{C^{(\beta)}_{11}(t)}{T^{2}_{1}}
= \left\{
\begin{array}{l}
\left[ \frac{3}{(1 + 2T_{1} t)^{3}} -
\frac{b_{1,2}(t)}{(1 + T_{1} t)^{4}} \right] \mathrm{e}^{-\gamma t}
\, \, \, \, \, \mbox{for}\, {\beta=1}, \\  \\
\left[ \frac{2}{(1 + T_{1} t)^{4}} -
\frac{2^{6} \, b_{2,2}(t)}{(2 + T_{1} t)^{6}} \right] \mathrm{e}^{-\gamma t}
\, \, \, \, \, \, \, \mbox{for}\, {\beta=2}, \\   \\
\left[ \frac{6}{(1 + T_{1} t)^{6}} -
\frac{2^{12} \, b_{4,2}(t)}{(2 + T_{1} t)^{10}} \right]
\mathrm{e}^{-2\gamma t} \, \, \, \mbox{for}\, {\beta=4},
\end{array} \right.
\end{equation}
where $b_{\beta,2}(t)$ is the two-level form factor~\cite{Guhr1998} and $T_{1}$
is the coupling strength, which is also extracted from the experiment via
$T_1=1-|\langle S_{11}\rangle|^2$ with the average $\langle S_{11} \rangle$
taken over the frequency.

In Fig.~\ref{fig:Parameters2chAB1B2TA} we show the autocorrelation function
$C^{(\beta)}_{11}(t)$ of the experimental data. The best fit yields $T_{1}=0.98$
and $\gamma=1.9$, for $\beta=1$, $T_{1}=0.96$ and $\gamma=0.5$, for $\beta=2$,
and $T_{1}=0.97$ and $\gamma=0.2$, for $\beta=4$, and they are plotted as dashed
lines. As expected the coupling parameters are almost the same for the three
symmetries but the absorption parameter is significantly different from one
symmetry to another. In particular, we notice that the value of $\gamma$ for
$\beta=2$ is almost twice the value for $\beta=4$. This may be due to the
interplay between reflection and absorption~\cite{Baez2008}, i.e., the higher the
reflection the smaller the absorption, and also due to the fact that $\gamma$ is given in units of $\Delta$ which is not the same for all graphs. This is the situation of the $\beta=4$ case which presents twice the reflection than that of the $\beta=2$ case (two subgraphs).
Also, the circulators introduce more reflections for $\beta=2$ and 4 in
comparison with the $\beta=1$ case with no circulators. The parameters $T_1$
and $\gamma$ are used in Eq.~(\ref{eq:Sabs}), from which we obtain $T_{31}$
and $T_{32}$, and finally compute $f$. The results are shown in
Fig.~\ref{fig:pfAB1B2B4} (lower panels) as dashed lines. A good agreement with
the experimental distribution is observed. For the symplectic case the agreement
between experiment and theory is good even without the correction due to
absorption and imperfect coupling; since $\gamma$ is relatively small,
$p_{4}(f)$ depends only weakly on the port couplings which are almost
perfect.

To conclude, we used three-terminal chaotic microwave graphs to measure the
different transmissions to extract the quantity $f$, that accounts for the
voltage drop in an equivalent quantum device and exhibits its nonlocal effects.
We successfully described the experimentally obtained distribution $p(f)$
providing analytical expressions for the ideal case, for the three symmetry
classes. Deviations from the ideal case for the orthogonal and unitary
symmetries are due to the presence of dissipation and imperfect coupling between
the graph and the junctions. Surprisingly, the dissipation and the coupling
strength do not erase the fingerprints of the symplectic symmetry. Our
experimental realizations are the first experiments of transport in
three-terminal systems for the three symmetry classes and the first experiment
with symplectic symmetry. We expect that our results motivate further studies
for a successful explanation of robustness of symplectic symmetry for imperfect
couplings and higher dissipation.

\acknowledgments

AMM-A acknowledges financial support from PRODEP under the project
DSA/103.5/16/11850. This work was partially supported by Fondo Institucional
PIFCA (Grant No. BUAP-CA-169), and CONACyT (Grants No. CB-2013/220624 and No.
CB-2016/285776). The experiments were funded by the Deutsche
Forschungsgemeinschaft via the individual grants STO 157/16-1 and KU 1525/3-1
including a short-term visit of AMM-A in Marburg.

%====================================================================++
%\bibliographystyle{apsrev-no-url}

%\bibliography{thispaper}

%====================================================================++

\end{document}